\documentclass[a4paper]{amsart}
\usepackage{amssymb,amsmath}
\usepackage{latexsym}
\usepackage{graphicx}
\usepackage{geometry}                
\usepackage{color}
\usepackage{xcolor}
\usepackage{pdfsync}
\usepackage{fancyhdr}
\usepackage[applemac]{inputenc}
\usepackage[breaklinks,colorlinks,backref]{hyperref}
\hypersetup{
    colorlinks, 
    linktoc=all, 
    linkcolor=red,  
  citecolor=hyptxt,
  urlcolor=blue}
  \hypersetup{
  citebordercolor=,
  filebordercolor=green,
  linkbordercolor=blue
}
\definecolor{hyptxt}{rgb}{0.7, 0.4, 0.9}
\usepackage{bibtopic}

\newcommand{\ov}{\overline}
\def\h{{\mathcal H }}

\def\R{\mathbb{R}}
\def\N{\mathbb{N}}
\def\C{\mathbb{C}}

\def\ud{\mathrm{d}}
\def\rg{\rangle}
\def\lg{\langle}
\def\adg{a^{\dag}}
\def\SN{\mathbb{S}}

\begin{document}

\title[Quantizations from (P)OVM's]{Quantizations from (P)OVM's}

\author{
H. Bergeron, 
E.M.F.  Curado,  
J.P. Gazeau, 
 Ligia M.C.S. Rodrigues}
\address{Univ Paris-Sud, ISMO, UMR 8214, 91405 Orsay, France}\email{herve.bergeron@u-psud.fr} 
\address{Centro Brasileiro de Pesquisas Fisicas, 
 Instituto Nacional de Ci\^encia e Tecnologia - Sistemas Complexos
Rua Xavier Sigaud 150, 22290-180 - Rio de Janeiro, RJ, Brazil} \email{evaldo@cbpf.br,jpgazeau@cbpf.fr, ligia@cbpf.br}
\address{APC, UMR 7164,
Univ Paris  Diderot, Sorbonne Paris Cit\'e, 
75205 Paris, France 
}\email{gazeau@apc.univ-paris7.fr}

\maketitle
\tableofcontents

\date{\today}

\begin{abstract}
We  explain the powerful role that  operator-valued measures  can play in quantizing any set equipped with a measure, for instance a group (resp. group coset) with its invariant (resp. quasi-invariant) measure.  Coherent state quantization is a particular case. Such integral quantizations are illustrated with two examples based on the Weyl-Heisenberg group and on the affine group respectively. An interesting application of the affine quantization in quantum cosmology is mentioned, and we sketch a construction of new coherent states for the hydrogen atom. 
\end{abstract}

\section{What is really quantization?}

The word ``quantization" is widely used in Physics, in Mathematics and in Signal Analysis. While the meaning  of quantization is perfectly clear to signal practitioners, the word  singularly lacks precision in  the two former domains, as was stated in the following quotation:
  \begin{quote} 
\textit{ First quantization is a mystery. It is the attempt to get from a classical description of a physical system to a quantum description of the ``same" system. Now it doesn't seem to be true that God created a classical universe on the first day and then quantized it on the second day. So it's unnatural to try to get from classical to quantum mechanics. Nonetheless we are inclined to do so since we understand classical mechanics better. So we'd like to find a way to start with a classical mechanics problem - that is, a phase space and a Hamiltonian function on it - and cook up a quantum mechanics problem - that is, a Hilbert space with a Hamiltonian operator on it. It has become clear that there is no utterly general systematic procedure for doing so } \cite{baez06} (see also the recent Todorov's review \cite{todorov12}). 
 \end{quote}
 
 In a nutshell, we might claim that ``\textit{ One quantizes things one does not really know to obtain things most of which one is unable to measure}". 

Let us recall the basic procedure, which rests upon the canonical commutation rules (ccr). Starting  from a $n$-dimensional phase space or symplectic manifold, it is summarized by the map ($n=2$):
\begin{align*}
 (q,p)\, ,  &\quad  \{q,p\}=1 \mapsto (Q,P)\,, \quad [Q,P] = i \hbar I\, ,\\
f(q,p) &\mapsto f(Q,P) \mapsto (\mathrm{Sym}f)(Q,P)\, .
\end{align*}
But then what about singular $f$, e.g. the phase $\arctan(p/q)$? What about barriers or other  impassable boundaries? 
Even the elegant  Weyl-Wigner integral  quantization \cite{weyl28,grossmann76,daub11,daub12,daub13}, which we revisit in this paper, is subject to serious difficulties when the geometry of the phase space is not  euclidean.

Let us propose here a  definition resting on three minimal requirements, linearity, identity, self-adjointness. More precisely, quantization is
\begin{enumerate} 
  \item[(i)] a \underline{linear} map $$\mathfrak{Q}: \mathcal{C}(X) \mapsto \mathcal{A}(\h)$$
where  $\mathcal{C}(X) $ is a vector space  of complex-valued functions $f(x)$ on a set $X$ and 
 $\mathcal{A}(\h)$ is a vector space  of linear operators $$\mathfrak{Q}(f) \equiv A_f$$  in some complex Hilbert space $\h$ such that
\item[(ii)]  $f=1$ is mapped to the identity operator $I$ on $\h$,
 \item[(iii)] a real  $f$  is mapped to an (essentially) self-adjoint operator $A_f$ in  $\h$.
\end{enumerate}
One is free to add  further requirements  on $X$ and $\mathcal{C}(X)$, e.g.,  measure,  topology,  manifold, closure under algebraic operations ...; to add
 physical  interpretation about measurement of spectra of  classical $f\in \mathcal{C}(X)$ or  quantum $\mathcal{A}(\h)$ to which are given the status of   \textit{observables}; to add dynamical considerations as time evolution; finally to add the  requirement of  unambiguous classical limit of the quantum physical quantities, the limit operation being associated with a change of scale.

\section{Integral quantization\cite{bergaz13,aagbook13,gazeaubook09}}

\subsection{Integral quantization: general setting}
We start from a 
 measure space $(X,\nu)$ where $X$ is the set of things we ``do not know well''. 
Suppose we were able to build  an $X$-labelled  family of bounded operators,
\begin{equation}
\label{xfamop}
 X\ni x \mapsto {\sf M}(x) \in \mathcal{L}(\h)\, ,  
\end{equation}
 on  Hilbert space $\h$,  resolving the identity $I$:
 \begin{equation}
\label{Psolveun}
\int_{X}\, {\sf M}(x)\, \ud\nu(x) = I\, , \quad \mbox{in a weak sense.}
\end{equation}
If the ${\sf M}(x)$'s are positive and unit trace, we will use the standard notation for density matrices :
$${\sf M}(x) \equiv \rho(x)\,. $$  
In this case, if $X$ is   equipped with a suitable topology, then  the map  $$\mathcal{B}(X)\ni \Delta \mapsto \int_{\Delta} \rho(x) \ud\nu(x)$$
may define a normalized positive operator-valued measure (POVM) $\mathfrak{m}$ on the $\sigma$-algebra $\mathcal{B}(X)$ of Borel sets
\begin{equation*}
\mathcal{B}(X) \ni \Delta \mapsto \mathfrak{m}(\Delta) = \int_{\Delta} \rho(x)\, \ud\nu(x)\,. 
\end{equation*}

Then the integral quantization of complex-valued functions $f(x)$ on $X$ is  the linear map:
 \begin{equation}
\label{formquant}
f\mapsto A_f = \int_{X} \, f(x)\, {\sf M}(x) \, \ud\nu(x)\, .
\end{equation}
If this map makes sense, the operator $A_f$ in $\h$ has to be understood  as the sesquilinear form, 
\begin{equation}
B_f(\psi_1,\psi_2)= \int_{X}f(x)\,  \lg \psi_1|{\sf M}(x)|\psi_2\rg \, \ud\nu(x)\,,
\end{equation}
defined on a dense subspace of $\h$. 
Note that  if $f$ is real and at least semi-bounded, the Friedrich's extension \cite{reedsimon2} of $B_f$ univocally defines a self-adjoint operator. 
If $f$ is not semi-bounded,  there is no natural choice of a self-adjoint operator associated with $B_f$ (see for instance  \cite{bergasiyou10,bersiyou12}).
A first observation about the above construction is its probabilistic content  
when ${\sf M}(x)= \rho(x)$.  Pick  another (or the  same)  family of positive unit trace operators $X\ni x \mapsto \widetilde{\rho}(x) \in \mathcal{L}^+(\h)$. Then we  return  to the classical world through the construction of the so-called lower (Lieb) or covariant (Berezin) symbol:
 \begin{equation}
\label{pbformquant}
A_f \mapsto \check{f}(x) := \int_{X} f(x')\,\mathrm{tr}(\widetilde{\rho}(x)\rho(x')) \,  \ud\nu(x')\, , \ \mbox{``lower symbol''}\, , 
\end{equation}
provided the integral be defined. Due to the fact that $x'\mapsto  \mathrm{tr}(\widetilde{\rho}(x)\rho(x'))$ is also a probability distribution, the map $f\mapsto \check{f}$ should be  viewed as a local  averaging (a ``blurring'') of the original $f$. It is also a generalization of the so-called Bargmann-Segal transform (see for instance \cite{stenzel94,hallmit05}). Quantization issues, e.g.  spectral properties of  $A_f$, may be derived from functional properties of  $\check{f}$.

Moreover, equipped with such a map, a classical limit condition can be considered in the following sense: given a scale parameter $\epsilon$ and a distance $d(f,\check{f})$:
\begin{equation}
\label{classlim1}
d(f,\check{f})\to 0 \quad \mbox{as}\quad \epsilon \to 0\,. 
\end{equation}
Another interesting aspect of the integral quantization map is the granted possibility of quantizing constraints: suppose that $(X,\nu)$ is  a smooth $n$-dim. manifold  on which is defined space $\mathcal{D}'(X)$ of distributions  as the topological dual of compactly supported $n$-forms on $X$ \cite{grosser08}.  Some of these distributions, e.g.  $\delta(u(x))$, express geometrical constraints. Extending the map $f\mapsto A_f$ to these objects yields the quantum version $A_{\delta(u(x))}$ of these constraints. 
There exists a different starting point, more in Dirac's spirit \cite{dirac64} and used in (Loop) Quantum Gravity and Quantum Cosmology. It would consist in determining the kernel of the operator  $A_u$ issued from integral quantization $u\mapsto A_u$. 
 Both methods are obviously not \emph{mathematically} equivalent, except for a few cases. They are possibly \emph{physically} equivalent, i.e. indistinguishable in terms of measurement.

\subsection{Covariant integral quantization based on  Unitary Irreducible Representation}

 Let $G$ be a Lie group with left Haar measure $\ud\mu(g)$, and let $g\mapsto U(g)$ be a unitary irreducible representation (UIR) of $G$ in a Hilbert space $\h$. 
Let ${\sf M}$ a bounded operator on $\h$. Suppose that the  operator 
\begin{equation}
\label{intgrR}
R:= \int_G  \, {\sf M}(g)\,\ud\mu(g) \, , \quad  {\sf M}(g):= U(g)\, {\sf M}\, U^{\dag}(g)\, , 
\end{equation}
is defined in a weak sense. From the left invariance of $\ud\mu(g)$ we have  $U(g_0)\,R\, U^{\dag}(g_0)= \int_G \ud\mu(g) \, {\sf M}(g_0g) = R$ and so $R$ commutes with all operators $U(g)$, $g\in G$.  Thus, from Schur's Lemma,  $R= c_{ {\sf M}}I$ with
\begin{equation}
\label{calcrho}
c_{ {\sf M}} = \int_G  \, \mathrm{tr}\left(\rho_0\, {\sf  M}(g)\right)\, \ud\mu(g)\, ,
\end{equation}
where the unit trace positive operator $\rho_0$ is  chosen in order to make the integral convergent. 
The resolution of the identity follows:
\begin{equation}
\label{Resunityrho}
\int_G \, {\sf M}(g) \,\ud \nu(g) = I\,, \quad \ud \nu(g):= \ud\mu(g)/c_{ {\sf M}}\, . 
\end{equation}
If UIR $U$ is square integrable with  $|\eta\rg$ being an admissible (unit) vector for $U$, i.e. $c(\eta):= \int_G \ud\mu(g) \, \vert \lg \eta| U(g)\eta\rg\vert^2 < \infty$, then 
the resolution of the identity is obeyed by \textit{coherent states} for $G$: $|\eta_g\rg= U(g)|\eta\rg$, i.e. by $\rho(g)= U(g)\,\rho \,U^{\dag}(g)$, where  $\rho= |\eta\rg\lg\eta|$. (A nice and meaningful  example of group having square integrable UIR's is the affine group, see Section \ref{affquant}). 
 This allows a \textit{covariant} integral  quantization of complex-valued functions on the group $f\mapsto A_f = \int_G\,  {\rho (g)}\,f(g)\, \ud \nu(g) $ :
\begin{equation}
\label{covintquant}
U(g) A_f U^{\dag}(g) = A_{U(g)f}\, ,  
\end{equation}
where $(U(g)f)(g'):= f(g^{-1}g')$ (regular representation if $f\in L^2(G,\ud\mu(g))$).
The Bargmann-Berezin-Segal or ``heat kernel'' transform on $G$ is then defined as $\check{f}(g) := \int_{G}\, \mathrm{tr}(\rho(g)\,\rho(g')) \, f(g')\,\ud\nu(g')$.

In the absence of square-integrability of $U$ over $G$, there may exist  a definition  of square-integrability and related  covariant coherent states with respect to a left coset manifold $X= G/H$, with $H$ a closed subgroup of $G$,  equipped with a quasi-invariant measure $\nu$. The most known example is the Weyl-Heisenberg group (see below). For more details and  examples see \cite{aagbook13}.

\section{Weyl-Heisenberg covariant integral quantization(s)}
\label{WHquant}
\subsection*{W-H quantization(s)}
We first recall the basic material and notations for the Weyl-Heisenberg (W-H) algebra and its Fock or number representation
 Let $\mathcal{H}$ be a separable (complex) Hilbert space  with orthonormal basis $e_0,e_1,\dots, e_n \equiv |e_n \rangle, \dots$. 
 Lowering and raising operators $a$ and $a^\dag$ are defined by their action on the basis:
\begin{equation}
\label{lowrais}
  a\, |e_n\rg  = \sqrt{n} |e_{n-1}\rg\, , \quad  a|e_0\rg = 0 \, , \quad 
   a^{\dag} \,|e_n\rg  = \sqrt{n +1} |e_{n+1}\rg\, ,
\end{equation}  
and the triplet  $\{a,\adg, I\}$ generate the Weyl-Heisenberg algebra  characterized by the canonical commutation rule
  \begin{equation}
\label{ccr}
  [a,\adg]= \mbox{I}\,.
\end{equation}
 There results from \eqref{lowrais} that the  number operator $N:= \adg a$ is diagonal with spectrum $\N$, $N|e_n\rg = n |e_n\rg$. It is well known that there exists an essentially unique UIR of the W-H algebra or group, at the root of quantum mechanics. Its square integrability holds with respect to the center $C\sim \R$ of the W-H group, and the measure space which has to be considered here is the euclidean or complex plane $X = G_{\mathrm{WH}}/C\sim \C$ with measure $\ud^2z/\pi$.
 To each $z\in \C$ corresponds the  (unitary) displacement operator  $D(z)$ :
\begin{equation}
\label{dispop}
\C \ni z \mapsto D(z) = e^{z\adg -\bar z a}\,, \quad D(-z) = (D(z))^{-1} = D(z)^{\dag}\,  . 
\end{equation}
The ccr \eqref{ccr} or QM non commutativity is encoded  by the addition formula
\begin{equation}
\label{adddisp}
D(z)D(z') = e^{\frac{1}{2}(z\bar{z'} -\bar{z} z')}D(z+z') \,.
\end{equation}
The standard (i.e., Schr\"odinger-Klauder-Glauber-Sudarshan) CS \cite{perelomovbook86} are then obtained by
\begin{equation}
\label{stCS}
|z\rg = D(z)|e_0\rg \,.
\end{equation}
Let $\varpi(z) $ be a function on the complex plane obeying $\varpi(0) = 1$ and chosen in such a way that 
 the operator-valued integral
\begin{equation*}
\label{opMvarpi}
 \int_{\mathbb{C}} D(z)\, \varpi(z) \frac{\ud^2 z}{\pi}:={\sf M}
\end{equation*}
defines (in a weak sense) a bounded operator ${\sf M}$ on $\h$.
Then the family of displaced ${\sf M}(z):= D(z) {\sf M}D(z)^\dagger$  under the unitary action $D(z)$ resolves the identity 
\begin{equation}
\label{residMz}
\int_{\mathbb{C}} \, {\sf M}(z) \,\frac{\ud^2 z}{\pi}= I\, . 
\end{equation}
This is a direct consequence of  $D(z) D(z') D(z)^\dagger = e^{z\ov{z}' -\ov{z} z'} D(z')$,  of
$
\int_{\mathbb{C}} e^{ z \bar \xi -\bar z \xi} \,  \frac{\ud^2 \xi}{\pi} = \pi \delta^{2}(z)\, ,
$
and of  $\varpi(0) = 1$ with $D(0)= I$.

The resulting quantization map is given by
\begin{equation}
\label{eqquantvarpi}
f\mapsto A_f = \int_{\mathbb{C}}  f(z) \, {\sf M}(z) \,  \frac{\ud^2 z}{\pi} = \int_{\mathbb{C}}  \hat{f}(-z)\,  D(z)\,  \varpi(z) \frac{\ud^2 z}{\pi}  \, , 
\end{equation}
where is involved the symplectic Fourier transform
$\hat{f}(z)=\int_{\mathbb{C}} f(\xi)\,e^{ z \bar \xi -\bar z \xi} \,  \frac{\ud^2 \xi}{\pi}$.
The covariance with respect to the translations reads as 
\begin{equation}
\label{covquant}
A_{f(z-z_0)} = D(z_0) A_{f(z)} D(z_0)^\dagger\, .
\end{equation}
Note the properties of $A_f$ resulting from properties of the weight function: 
\begin{align*}
 A_{f(-z)} &= {\sf P} A_{f(z)} {\sf P}, \forall \,f\  \  \iff \   \varpi(z)=\varpi(-z), \,\forall \,z\, ,\\
 A_{\overline{f(z)}}& = A_{f(z)}^\dagger, \forall \,f   \ \iff \   \overline{\varpi(-z)}=\varpi(z), \, \forall \,z\, ,
  \end{align*}
where ${\sf P} = \sum_{n=0}^{\infty}(-1)^n|e_n\rg\lg e_n|$ is the parity operator.

Now, the ccr is almost always the rule when $\varpi$ is chosen \emph{ real } and \emph{even} . 
Indeed we have in this case
\begin{equation}
\label{regquant1}
A_{z} = a\, , \quad  A_{\overline{f(z)}} = A_{f(z)}^\dagger\, , 
\end{equation}
or equivalently, with $z= (q+ip)/\sqrt 2$, 
\begin{equation}
\label{regquantqp1}
A_q = \frac{a + \adg}{\sqrt{2}} := Q \, , \, A_p= \frac{a-\adg}{i \sqrt{2}} := P\,, \ [Q,P]= iI\, . 
\end{equation}
Morover,  iff $\vert \varpi(z) \vert=1$,
\begin{equation}
\mathrm{tr} (A_f^\dagger A_f) = \int_\C \vert f(z)\vert^2 \frac{\ud^2z}{\pi}\, , 
\end{equation}
which means that the map $L^2(\C,d^2z/\pi)\ni f \mapsto A_f\in \h_{\mathrm{Hilbert-Schmidt}}$ is invertible through a trace formula.
\subsection*{Cahill-Glauber weight}
For instance let us choose as a weight function the exponential
\begin{equation}
\label{exexpos}
\varpi_s(z) = e^{s \vert z\vert^2/2}\, , \quad \mathrm{Re}\; s<1 \,.
\end{equation}
This yields a diagonal ${\sf M}\equiv {\sf M}_s$ with
\begin{equation}
\label{diagMs}
\lg e_n|{\sf M}_s|e_n\rg = \frac{2}{1-s}\,\left(\frac{s+1}{s-1}\right)^n\, , 
\end{equation}
and so
\begin{equation}
\label{defMs}
{\sf M}_s= \int_{\mathbb{C}} D(z) \, \varpi_s(z)\,\frac{{\ud}^2z}{\pi }= \frac{2}{1-s} \exp \left\lbrack \ln \left(\dfrac{s+1}{s-1}\right)\, a^\dag a \right\rbrack\,.
\end{equation}

The value $s=-1$ corresponds to the CS  (anti-normal) quantization, since 
\begin{equation}
{\sf M}= \lim_{s\to -1} \dfrac{2}{1-s} \exp \left( \ln \dfrac{s+1}{s-1} a^\dag a \right) = |e_0\rg\lg e_0|\, , 
\end{equation}
and so 
\begin{equation}
\label{csquants-1}
A_f = \int_{\mathbb{C}}  \, D(z){\sf M}D(z)^{\dagger} \, f(z) \,\frac{\ud^2 z}{\pi}= \int_{\mathbb{C}} \,  |z\rg\lg z| \, f(z)  \,\frac{\ud^2 z}{\pi}\, .
\end{equation}
The choice $s=0$ implies $
\sf M= 2\sf P
$ and corresponds to the Wigner-Weyl quantization. Then
\begin{equation}
\label{wigweylquant}
A_f = \int_{\mathbb{C}} \,  D(z) \,2{\sf P}\, D(z)^{\dagger} \, f(z) \,\frac{\ud^2 z}{\pi}\, .
\end{equation}
The case $s=1$ is the normal quantization obtained in an asymptotic sense. 

The function \eqref{exexpos} was originally introduced  by Cahill and Glauber \cite{cahillglauber69,cahillglauber69_2}
in view of discussing the problem of expanding an arbitrary operator as an ordered power series in  $a$ and $\adg$, a typical question encountered in quantum field theory, specially in quantum optics. Actually, they were not  interested in the question of quantization itself. Now, a very interesting feature of  \eqref{exexpos} is that the operator ${\sf M}_s$ is  positive unit trace class for $s \leq -1$ (and only trace class if $\mathrm{Re}\;s<0$), i.e.,  is a density operator, ${\sf M}_s\equiv \rho_s$. Therefore, in the range $s \leq -1$ the corresponding quantization has a consistent  probabilistic content in the sense that the operator-valued measure 
 \begin{equation*}
\label{spovs}
\C \supset \Delta \mapsto  \int_{\Delta\in \mathcal{B}(\C)}  D(z){\sf M}_sD(z)^{\dag}\,\dfrac{\ud^2 z}{\pi}  
\end{equation*}
is a \underline{POVM}. Moreover, a Boltzmann-Planck expression  with temperature $T$ can be  associated with  this lack of knowledge of the classical $X=\C$, say a kind of noise temperature as we have in electronics. Given an elementary quantum energy, say $\hbar \omega$ and with the temperature $T$-dependent  
$s = - \coth\dfrac{\hbar \omega}{2k_B T}
$
the density operator quantization is Boltzmann-Planck
 \begin{equation}
\label{plboltrho}
\rho_s= \left( 1- e^{-\tfrac{\hbar \omega }{k_B T}}\right)\sum_{n=0}^{\infty} e^{-\tfrac{n\hbar \omega }{k_B T}}|e_n \rg\lg e_n|\,. 
\end{equation} 
Thus,  the temperature-dependent operators $\rho_s(z) = D(z) \, \rho_s\, D(z)^{\dag}$ defines a  Weyl-Heisenberg covariant  family of POVM's on the phase space $\C$, the null temperature limit case being the  POVM built from standard CS. We notice that the Weyl-Wigner integral quantization ($s=0$) and its associated Wigner function are out of the scope of this thermodynamic consideration. 

\subsection*{Quantum harmonic oscillator according to $\varpi$}
We saw that any real even $\varpi$ defining a bounded operator through \eqref{opMvarpi}  yields the ccr for the quantization of the canonical pairs $(z,\bar z)$ or, equivalently $(q,p)$. Actually it yields also the correct  energy spectrum for the harmonic oscillator. Indeed, we have
\begin{align*}
     A_{q^2} &= Q^2 - \left.\partial_{z}\partial_{\bar z}\varpi\right\vert_{z=0} + \frac{1}{2} \left( \left.\partial_{z}^2 \varpi\right\vert_{z=0}+ \left.\partial_{\bar z}^2 \varpi\right\vert_{z=0}\right)\, ,   \\
      A_{p^2}&= P^2 - \left.\partial_{z}\partial_{\bar z}\varpi\right\vert_{z=0} - \frac{1}{2} \left( \left.\partial_{z}^2 \varpi\right\vert_{z=0}+ \left.\partial_{\bar z}^2 \varpi\right\vert_{z=0}\right)\, ,
\end{align*}
 and so
 \begin{equation}
\label{enrHO}
A_{\vert z \vert^2} \equiv A_J=  \adg a + \frac{1}{2}  -  \left.\partial_{z}\partial_{\bar z}\varpi\right\vert_{z=0}\,.
\end{equation}
where $\vert z \vert^2 (= J)$ is the energy (or action variable) for the H.O. 
The difference between the ground state energy
 $E_0= 1/2- \left.\partial_{z}\partial_{\bar z}\varpi\right\vert_{z=0}$, and the minimum of the quantum potential energy  $E_m=[\min(A_{q^2}) + \min(A_{p^2})]/2 = - \left.\partial_{z}\partial_{\bar z}\varpi\right\vert_{z=0}$ is  $E_0-E_m=1/2$. So it is the  (experimentally verified, see for instance \cite{herzberg1989}) half quantum (in appropriate units),  independently of the particular quantization which has been chosen. 
In the exponential Cahill-Glauber case $\varpi_s(z) = e^{s\vert z \vert^2/2}$ the above operators reduce to 
\begin{equation*}
\label{quantosc2}
A_{\vert z \vert^2} = \adg a + \frac{1-s}{2}\,, \quad A_{q^2}=Q^2-\frac{s}{2} \,,  \quad A_{p^2}=P^2 - \frac{s}{2} \,.
\end{equation*}
More details and relevant references on this question are found in  \cite{bergayou13} where it is  proven  that these constant shifts in energy are inaccessible to measurement. 

\subsection*{ Weyl-Heisenberg integral quantization with  action-angle variables}

With  $z = \sqrt J\, e^{i\gamma}$ in action-angle $(J,\gamma)$ notations for the harmonic oscillator, the quantization of  $f(J, \gamma)$, $2\pi$-periodic in $\gamma$,  yields formally 
\begin{equation}
\label{aaquanta}
A_{f} = \int_0^{+\infty}\ud J \int_0^{2\pi}\frac{\ud\gamma}{2\pi} f(J,\gamma){\sf M}\left(\sqrt{J}e^{i\gamma}\right)\,. 
\end{equation}
We define the unitary representation $\theta \mapsto U_{\mathbb{T}}(\theta)$ of the unit circle  $\SN^1$ on the Hilbert space $\mathcal{H}$ as  $U_{\mathbb{T}}(\theta)|e_n\rg = e^{i (n + \nu) \theta}|e_n\rg$, where $\nu$ is arbitrary real. If the operator ${\sf M}$ is diagonal, then one  has  the angular covariance property:
\begin{equation}
\label{covquantaa}
U_{\mathbb{T}}(\theta)A_f U_{\mathbb{T}}(-\theta)= A_{T(\theta)f}\, , \quad T(\theta)f(J,\gamma) = f(J, \gamma -\theta)\, . 
\end{equation}
 In particular, let us quantize with coherent states, ${\sf M}(z)= \rho_{-1}(z)= |z\rg\lg z|$, the discontinuous $2\pi$-periodic angle function $\gimel(\gamma) = \gamma$ for $\gamma \in [0, 2\pi)$.  In terms of the action-angle variables   these CS  read as  
 \begin{equation}
|z\rg \equiv |J,\gamma\rg = \sum_n \sqrt{p_n(J)} e^{in\gamma} |e_n\rg\, , 
\end{equation}
where $ n \mapsto p_n(J) = e^{-J} J^n/n!$ is the Poisson distribution.    Since the angle function is real and bounded,  its quantum counterpart $A_{\gimel}$  is a bounded self-adjoint operator, and it is covariant in the above sense. 
 In the basis $|e_n\rg$, it is given by  the infinite matrix:
\begin{equation}
\label{scsphaseop}    
A_{\gimel}= \pi\,  1_{{\mathcal H}} + i \, \sum_{n\neq n'}\frac{\Gamma\left( \frac{n + n'}{2}+1\right)}{\sqrt{n!n'!}}\, \frac{1}{n'-n}\, |e_n\rg\lg e_{n'}|\, .
\end{equation}
This quantum angle has spectral measure with support $[0,2\pi]$. Of course, plenty of similar quantum angles are made possible with  that freedom we have in choosing the weight function. It is an interesting question to be considered from different viewpoints, particularly from measurement viewpoints based on POVM,  as they are described in \cite{bushgrablahti95,royer96}.

\section{Affine quantization}
\label{affquant}
\subsection*{General setting}
This is the second basic illuminating example of the method.  Like the above complex plane and abelian group $X=\C$, is viewed as the phase space for the motion of a particle on the line,  the half-plane is a group which can be viewed as the phase space for the motion of a particle on the half-line. Let us be more precise. 
Our measure space $(X,\nu)$ is the upper half-plane $X\equiv  \Pi_+:= \{(q,p)\, |\, p\in \mathbb{R}\, , \, q> 0\}$ equipped with the left invariant measure $\ud q \ud p$. 
Equipped with the multiplication 
$(q,p)(q_0,p_0)=(qq_0, p_0/q+p), ~q\in \mathbb{R}^{\ast}_+, ~p\in \mathbb{R}$, 
$\Pi_+$ is viewed as the  affine group Aff$_+(\R)$ of the real line.
 Aff$_+(\R)$ has
two non-equivalent  UIR, $U_{\pm}$ \cite{gelnai47,aslaklauder68}. Both are square integrable and this is the rationale backing the  \textit{continuous wavelet analysis} {\cite{grosmor84,grosmorpaul85,grosmorpaul86,aagbook13}. 
The UIR $U_+\equiv U$ is carried on by  Hilbert space $\mathcal{H} = L^2(\mathbb{R}^{\ast}_+,  dx)$:
\begin{equation}
\label{affrep+}
U(q, p) \psi(x) = (e^{i px}/\sqrt{q}) \psi( x/q)\, .
\end{equation}
As we did for the Weyl-Heisenberg group, we pick a suitably localized weight function ${\sf w}(q,p)$ on the half-plane such that the integral 
\begin{equation}
\label{affbop}
\int_{ \Pi_+} U(q,p)\, {\sf w}(q,p)\, \ud q\, \ud p:= {\sf M}
\end{equation}
defines a bounded operator in a weak sense . Proceeding with the same construction as in \eqref{intgrR}, \eqref{calcrho} and \eqref{Resunityrho} yields the resolution of the identity on 
\begin{equation}
\label{affresw}
\int_{ \Pi_+} {\sf M}(q,p)\, \frac{\ud q\, \ud p}{c_{{\sf w},{\sf M}}}= I\, , \quad {\sf  M}(q,p)= U(q,p){\sf M}U^{\dag}(q,p)\, ,
\end{equation}
and the resulting covariant quantization based on the affine group:
\begin{equation}
\label{qaffw}
f\mapsto A_f = \int_{ \Pi_+} f(q,p)\, {\sf M}(q,p)\, \frac{\ud q\, \ud p}{c_{{\sf w},{\sf M}}}\,. 
\end{equation}
Due to the square-integrability of $U$, the simplest choice to be made is ${\sf M} = |\psi\rg\lg \psi | = \rho_0$ where the unit-norm state $\psi \in L^2(\mathbb{R}_+^\dagger, \ud x)\cap L^2(\mathbb{R}_+^\dagger, \ud x/x)$ (``fiducial vector" or ``wavelet'')
produces all  affine coherent states, i.e. wavelets,  defined as $| q, p \rangle\ = U(q,p) | \psi \rangle$. Resolution of the identity and resulting covariant quantization now read:
\begin{equation}
\label{affresid}
\int_{\Pi_+}  | q, p \rangle \langle q, p | \, \dfrac{\ud q \ud p}{2 \pi  c_{-1}}= I \, , \ 
c_{\gamma}:=\int_0^\infty  \vert\psi(x)\vert^2\, \frac{\ud x}{x^{2+\gamma}}\, ,
\end{equation}
\begin{equation}
\label{ affqpsi}
f \ \mapsto \ A_f = \int_{\Pi_+}   f(q, p) |q, p \rangle \langle q, p |\, \dfrac{\ud q \ud p}{2\pi  c_{-1}}\, . 
\end{equation}
(Had we chosen $U_-$ we would produce identical results. It is just a matter of taste between negative or positive half-line.) 
  The quantization is canonical  for $q$ and $p$, in the sense that the ccr gives $i$ times a constant:
\begin{equation}
\label{canaffquant}
A_p= P = -i\partial/\partial x\, ,\ A_{q^\beta} =({c}_{\beta-1}/c_{-1})\, Q^\beta\, , \ Qf(x) =  x f(x)\,. 
\end{equation}
Note the multiplicative factor of $Q^\beta$, absent in the W-H quantization: this is the price to pay for dealing with dilations. 
Now, the important point concerns the quantization of the kinetic energy:
\begin{equation}
\label{affqkinen}
  A_{p^2} = P^2 +  KQ^{-2} \, , \quad K=K(\psi)= \int_0^{\infty} (\psi'(u))^2\, u\, \frac{\ud u}{c_{-1}}\, . 
\end{equation}
Thus, due to the presence of repulsive ($\sim$ centrifugal) potential  this affine or 
 wavelet quantization prevents a quantum free particle moving on the positive line from reaching the origin.
We know (see \cite{reedsimon2}) that  the operator $P^2= -d^2/dx^2$ alone, in $L^2(\mathbb{R}^{\ast}_+,  \ud x)$, is not essentially self-adjoint whereas the above regularized operator, defined on the domain   $C_0^{\infty}(0, \infty)$  of smooth compactly supported functions,  is essentially self-adjoint for $K \geq 3/4$. Then quantum dynamics of the free motion is possible. 

Quantum states and their dynamics have phase space representation through lower symbols. With a  state  $| \phi \rangle$  is associated the probability distribution on the phase space:
$$
 \rho_\phi(q, p) = \dfrac{1}{2\pi  c_{-1}} |\langle q, p | \phi \rangle|^2\,  .
$$
With energy eigenstates at our disposal, we can compute the time evolution for any state and the  formula for the associated time-dependent probability distribution follows. 

The integral quantization based on the affine group has recently found interesting applications in quantum cosmology, where the singularity at zero volume of the universe is naturally regularized with such a scheme \cite{wcosmo1,wcosmo2}. 
Note that proceeding in quantum  theory with an ``affine'' quantization instead of the  Weyl-Heisenberg quantization  was already present in Klauder's  work devoted to the question of dealing with singularities in quantum gravity \cite{klauder11,fanuelzonetti13}. The procedure rests on the representation of the affine Lie algebra. In this sense, it remains closer to the canonical one and it is not of the integral type.

\subsection*{As a byproduct: affine CS for central potentials, e.g. Coulomb-Kepler}
We sketch now the content of a work in progress \cite{bercugaroH13} where we partly use the material presented in this paper. We know that the construction of coherent states for the hydrogen atom which would have as much rich properties as the standard CS have for the harmonic oscillator is still an open (if solvable!) problem. Several systems of CS have been proposed
since Schr\"odinger's original work, derived mainly from the groups SU(1,1)
and SO(4,2), although in a regularized sense (the latter group is the full dynamical group of the H-atom; the
former is the subgroup describing its radial motion only), but none of them
is fully convincing (see \cite{klauder96} for a list of references). The construction that we propose is based on the above affine CS for the radial part and on the Kowalski-Rembielinski-Hall-Mitchell CS for the angular part. Indeed, we expect that both have good localization properties in the phase space for the motion of a charged particle submitted to the Coulomb potential. 
We know that the   quantum hamiltonian $H = -\Delta +k/r^2 -g/r$ with domain $C_0^{\infty}(\R^3)$ in $L^2(\R^3, \ud^3 \mathbf{ r})$ 
is self-adjoint \cite{reedsimon2}. Using spherical coordinates $\mathbf{r} = (r,\hat{\mathbf{y}})$ , a state $\Psi(\mathbf{ r})$ in $L^2(\R^3, \ud^3 \mathbf{ r})$ factorizes as 
\begin{equation}
\label{factHat}
\Psi(\mathbf{ r}) = r \psi(r) \mathcal{Y}(\hat{\mathbf{y}})\, , 
\end{equation}
where  $\psi \in L^2(\R^+,\ud r)$ and $\mathcal{Y}\in L^2(\mathbb{S}^2, \ud \hat{\mathbf{y}})$, where $\hat{\mathbf{y}}$ is the SO(3) invariant measure on the 2-sphere. Use the material above with a suitable fiducial vector $\psi(r)$ yields affine CS for the radial part which are labeled by points $(q_r,p_r)$ of the upper half-plane:
\begin{equation*}
\mathcal{R}_{q_r,p_r}(r)  = \left(U(q_r,p_r)\psi\right)(r)\,. 
\end{equation*}
Now, the phase space for the motion on sphere $\mathbb{S}^2= \{\mathbf{x} \in \R^3\, , \, x^2 = \sum_k x_k^2= 1 \}$ is realized as the complexified sphere $\mathbb{S}_C^2$
\begin{equation}
\label{compsphere}
T^{\ast}({\SN}^2) \simeq {\SN}_c^2 = \{ \mathbf{a} = (a_1, a_2,a_3)  \in \C^{3} : a^2 = \sum_k a_k^2= 1 \}\,,
\end{equation}
with, in suitable units,  
\begin{equation}
\label{paraSc}
 \mathbf{a} = (\cosh{J})\, \mathbf{x}
+ i \frac{1}{J} (\sinh J)\, \mathbf{p}\, , \quad J= \Vert \mathbf{x}\wedge\mathbf{p}\Vert\,, \quad   \mathbf{x}\cdot  \mathbf{p} = 0\, . 
\end{equation}
The Kowalski-Rembieli\'nski 
 coherent states $| \zeta_{\mathbf{a}}\rg$ are   realized as elements of the Hilbert space $L^2(\mathbb{S}^2,\ud \hat{\mathbf{y}})$ as follows
\begin{equation}
\label{CSL2Sd}
\zeta_{\mathbf{a}}(\hat{\mathbf{y}}) = = (2 \pi)^{-1}
  \frac{e^{1/8}}{\sqrt{\pi }}\int_{\varOmega}^{\pi}\frac{\ud\phi}{\sqrt{\cos{\varOmega} - \cos{\phi}}}\, ,
  \end{equation}
which depends only on the complex angle $\varOmega= \varOmega(\mathbf{a},\hat{\mathbf{y}})$ defined through analytic continuation by
 \begin{equation}
\label{varomega}
\cos{\varOmega} = \mathbf{a}\cdot\hat{\mathbf{y}}= (\cosh{J})\, \mathbf{x}\cdot\hat{\mathbf{y}}
+ i \frac{1}{J}(\sinh{J})\, \mathbf{p}\cdot\hat{\mathbf{y}}\,. 
\end{equation}
They solve the identity in $L^2(\mathbb{S}^2_c, \ud \mu_h)$  with an explicit measure given in \cite{kowrem00,hallmit02}.
Finally, the coherent states for the particle read
\begin{equation}
\label{cshatom}
\mathcal{Z}_{q_r,p_r, \mathbf{a}}(\mathbf{ r}) = r\mathcal{R}_{q_r,p_r}(r) \, \zeta_{\mathbf{a}}(\hat{\mathbf{y}}) \, . 
\end{equation}

\section{Conclusion}
Beyond the  freedom (think to the analogy with Signal Analysis where different techniques are employed in a  complementary way) allowed by integral quantization,  the advantages of the method with regard to other quantization procedures in use are of four  types. 
 
 \begin{enumerate}
  \item[(i)] The minimal amount of constraints imposed to the classical objects to be quantized. 
  \item[(ii)] Once a choice of (positive) operator-valued measure has been made, which must be consistent with experiment,  there is no ambiguity in the issue, contrarily to other method(s) in use (think in particular of the ordering problem). To one classical object corresponds one and only one quantum object. Of course different choices are requested to be physically equivalent
  \item[(iii)] The method produces in essence a regularizing effect, at the exception of certain choices,  like the Weyl-Wigner integral quantization. 
  \item[(iv)] The method, through POVM choices, offers the possibility to keep a full probabilistic content. As a matter of fact,  the Weyl-Wigner integral quantization does not rest on a POVM. 
  \end{enumerate}
But what is the real meaning of that freedom granted to us in the choice of POVM or others?
Such a freedom is governed by our degree of confidence in localizing a pure classical state $(q,p)$ in phase space. The latter is usually viewed as an ideal continuous manifold where all points are physically accessible. As everybody knows, such a view is physically untenable ...
 However, and this is the paradoxical paradigm of contemporary physics, one needs such a  leibnizian mathematical ideality (\emph{natura non saltum facit}) to build a more realistic, though still highly mathematical, representation of the physical world.

\end{document}